\documentclass[graybox]{svmult}

\usepackage{mathptmx}       
\usepackage{helvet}         
\usepackage{courier}        
\usepackage{makeidx}         
\usepackage{graphicx}        
\usepackage{multicol}        
\usepackage[bottom]{footmisc}

\usepackage{amsfonts}
\usepackage{amsmath}
\usepackage{units}

\def\bestfrac#1#2{%
    \raise.2ex\hbox{$#1$}%
    \kern-.1em/\kern-.15em%
    \lower.25ex\hbox{$#2$}}

\makeindex             

\let\a=\alpha \let\b=\beta    \let\g=\gamma     \let\d=\delta     
             \let\l=\lambda
          \let\x=\xi                
          \let\ph=\varphi   
\let\ps=\psi   \let\o=\omega     
\let\G=\Gamma \let\D=\Delta           
             \let\Ps=\Psi
\let\O=\Omega

\def\cU{\mathcal{U}}
\def\cV{\mathcal{V}}
\def\cF{\mathcal{F}}

\def\cL{\mathcal{L}}
\def\cN{\mathcal{N}}
\def\cE{\mathcal{E}}
\def\cK{\mathcal{K}}
\def\cH{\mathcal{H}}

\def\ee{{\underline \varepsilon}}

  \def\v0{{\vec 0}}

\def\tl#1{{\tilde{#1}}}

\def\bC{\mathbb{C}}
\def\bR{\mathbb{R}}
\def\bN{\mathbb{N}}

\def\ph{\varphi}

\def\indic{\hbox{\raise-2pt \hbox{\indbf 1}}}

\let\dpr=\partial

\let\==\equiv

\let\io=\infty
\let\0=\noindent

\def\*{{\hfill\break\null\hfill\break}}
\def\bra#1{{\langle#1|}}
\def\ket#1{{|#1\rangle}}

\def\bmedia#1{{\bigl\langle#1\bigr\rangle}}

\def\ie{\hbox{\it i.e.\ }}

\def\tende#1{\,\vtop{\ialign{##\crcr\rightarrowfill\crcr
             \noalign{\kern-1pt\nointerlineskip}
             \hskip3.pt${\scriptstyle #1}$\hskip3.pt\crcr}}\,}
\def\otto{\,{\kern-1.truept\leftarrow\kern-5.truept\to\kern-1.truept}\,}

\def\Tr{{\rm Tr}}

\def\sqt[#1]#2{\root #1\of {#2}}

\def\Im{{\rm Im}\,}
\def\lis{\overline}

\newcommand{\di}{\textrm{d}}

\def\be{\begin{equation}}
\def\ee{\end{equation}}

\def\[#1\]{\begin{align}#1\end{align}}

\begin{document}

\title*{Analysis of fluctuations around non linear effective dynamics}

\author{Serena Cenatiempo}
\institute{Serena Cenatiempo \at Gran Sasso Science Institute, Viale Francesco Crispi 7, 67100 L'Aquila, Italy, \\ \email{serena.cenatiempo@gssi.infn.it}}

\maketitle

\abstract{
We consider the derivation of effective equations approximating the many-body quantum 
dynamics of a large system of $N$ bosons in three dimensions, interacting through a two-body potential $N^{3\beta -1} V(N^\beta x)$. For any $0\leq \b\leq 1$
well known results establish the trace norm convergence of the  k-particle reduced density matrices associated with the solution of the many-body Schr\"odinger equation towards products of solutions of a one-particle non linear Schr\"odinger equation, as $N \to \io$. In collaboration with C. Boccato and B. Schlein we studied fluctuations around the approximate non linear Schr\"odinger dynamics, obtaining for all $0<\beta<1$ a norm approximation of the evolution of an appropriate class of data on the Fock space. 
}

\section{Introduction} \label{sec:1}

The understanding of the properties of many body quantum systems is a challenging topic in quantum mechanics, the challenge being how to derive from the microscopic and fundamental description of the system those collective properties which are successfully exploited in condensed matter laboratories.  The analysis of the time evolution of quantum many particle systems and the derivation of effective descriptions in interesting limiting regimes nestle in this research line. From a mathematical physics perspective the main goals  in this field are on one hand to justify the use of effective equations, and on the other hand to clarify the limits of applicability of the effective theories. 

While we refer to \cite{Heidelberg} for an introduction on this topic, and a panorama on existing results and open problems in the context of bosonic and fermionic systems, we focus here on some recent results concerning the analysis of the time evolution of bosonic systems in three dimensions. 

A bosonic system of $N$ particles moving in three space dimensions can be described through a complex-valued wave function $\ps_N$ on the Hilbert space of permutation symmetric $L^2(\bR^{3N})$ wave functions
\begin{eqnarray}
L^2_s(\bR^{3N}) =   \{ && \ps_N \in L^2(\bR^{3N}):      \nonumber  \\
&& \ps_N(x_{\pi(1)}, \ldots, x_{\pi(N)})=\ps_N(x_1, \ldots, x_N) \; \text{for any permutation } \pi \in S_N \}   \nonumber 
\end{eqnarray}
with $\|\ps_N\|_2=1$. 
The evolution of an initial wave function $\ps_{N,0} \in L^2_s(\bR^{3N})$ is governed by the Schr\"odinger equation
\[ \label{Schr}
i \dpr_t \ps_{N,t} = H_N \ps_{N,t}\,,
\] 
where the subscript $t$ indicates the time dependence of $\ps_{N,t}$ and $H_N$ is a self adjoint operator on $L^2_s(\bR^{3N})$ known as Hamiltonian of the system. 
We will restrict our attention to Hamiltonians with two body interactions, and we will consider interactions scaling with the number of particles, as follows:
\begin{equation} \label{HN}
H_N^{(\b)} = \sum_{i=1}^N (-\D_{x_i}) + \sum_{i<j}^N N^{3\b -1} V(N^{\b}(x_i -x_j))\,,
\end{equation}
with $0 \leq \b \leq 1$ and $V$ a spherically symmetric interaction potential. We will be interested in situations where the number of particles $N$ is large. 

For $\b=0$ the Hamiltonian \eqref{HN} describes $N$ bosons interacting by a mean field potential $ N^{-1} V(x_i - x_j)$; this regime is a first approximation for the behaviour of dilute Bose gases and is characterized by very weak interactions for large $N$. 
A more accurate model for interactions among bosons in experiments on cold gases is given by the so called \emph{Gross-Pitaevskii regime}, which corresponds to the $\b=1$ case in \eqref{HN}.   In this regime the interaction scales as $N^{2} V(N(x_i -x_j))$, corresponding to a situation where there are strong and short range collisions.  While in the mean field regime, as we will see, correlations among particles can be neglected in order to obtain a (first) effective description of the system,  they play a crucial role in the Gross-Pitaevskii regime, due to the singularity of the potential.
Values of $\b$ between zero and one describe intermediate scalings between the mean field and Gross-Pitaevskii regimes, and therefore we may expect correlations to become more and more important as $\b$ approaches one.

We will be interested in studying the evolution under \eqref{HN} of a particular class of initial data, 
exhibiting {\it complete condensation}, meaning that there exists a one-particle wave function $\ph \in L^2(\bR^3)$ (the so called \emph{condensate wave function}) such that  the one-particle reduced density matrix associated to the many body wave function $\ps_{N,0}$ 
\begin{equation} 
 \g^{(1)}_{N,0} := N\Tr_{2 \ldots N} \ket{\ps_{N,0}}\bra{\ps_{N,0}} \,,
\end{equation}
 satisfies
\begin{equation} \label{BEC}
\Tr \Big| \,N^{-1}\g^{(1)}_{N,0} - \ket{\ph}\bra{\ph}\, \Big| \to 0\qquad \text{as } N \to \io\,,
\end{equation}
where $\ket{\ph}\bra{\ph}$ denotes the orthogonal projection onto $\ph$. From a physical point of view bosonic quantum states such that the one-particle reduced density matrix has an eigenvalue of order $N$ in the limit of large $N$ are models for \emph{Bose-Einstein condensates}, as realized in experiments on low density cold gases since 20 years~\cite{BECexperiments}.

We recall that the expectation of a bounded one particle observable $O^{(1)}$ on the many particle state described by $\ps_{N,0}$ is given by $\Tr( \g^{(1)}_{N,0} O^{(1)})$. Therefore whenever \eqref{BEC} occurs the knowledge of the condensate wave function is sufficient to determine the expectation of any bounded observable on the state described by $\ps_{N,0}$ in the limit $N \to \io$. Additionally, since  property \eqref{BEC} for bosonic systems also implies that for any $k =2,3, \ldots, N$ the $k$-particle reduced density matrix $\g^{(k)}_{N,0} = \binom{N}{k}\Tr_{k+1, \ldots, N}\ket{\ps_{N,0}}\bra{\ps_{N,0}}$  is given by a rank one projection onto $\ph^{\otimes k}$, we can also calculate the expectation of \emph{any} bounded $k$-particle observable in the limit $N \to \io$  .  

Now, let us start with an initial datum satisfying \eqref{BEC} and  let the system evolve with Hamiltonian \eqref{HN}. Due to the presence of the interaction we cannot expect \eqref{BEC} to hold at positive times. However one can show that this property remains approximately true in the limit of large $N$. Furthermore, one can derive an effective dynamics for the condensate wave function in the scaling regimes described by \eqref{HN}. More precisely one can prove (references will follow at the end of this section) that, for every family of initial data $\ps_{N,0} \in L^{2}(\bR^{3N})$ satisfying \eqref{BEC},  the one-particle reduced density matrix $\g^{(1)}_{N,t}$ corresponding to  the evolved state $\ps_{N,t} = e^{-i t H_N^{(\b)}} \ps_{N,0}$ (under suitable assumptions on the interacting potential)  satisfies
\begin{equation} \label{tracenormconvergence}
\Tr \big| N^{-1}\g^{(1)}_{N,t}  -  \ket{\ph_t} \bra{\ph_t}\big| \xrightarrow[N \to \io]{}  0\,,
\end{equation}
with $\ph_t$ solution of a non linear Schr\"odinger equation with initial datum $\ph_0=\ph$, whose precise form depends on $\b$. In particular $\ph_t$ satisfies: 
\begin{align}
 i \dpr_t \ph_t & = - \D \ph_t + (V \star |\ph_t|^2) \ph_t & \text{if }  \b=0\,,  \label{Hartree}\\
  i \dpr_t \ph_t & = - \D \ph_t + (\int V)\,  |\ph_t|^2 \ph_t & \text{if } 0<\b<1\,, \label{NLS} \\
   i \dpr_t \ph_t & = - \D \ph_t +8 \pi a_0 |\ph_t|^2 \ph_t &\text{if }  \b=1 \label{GP} \,.
\end{align}
The parameter  $a_0$ appearing in \eqref{GP} is the scattering length of the interaction $V$, \ie $8 \pi a_0= \int V(x)f(x)$ with $f(x)$  the solution of the zero energy scattering equation
\be \label{scattering}
\big(-\D + \frac 12 ) f =0\,,\qquad f(x) \xrightarrow[x \to \io]{} 1\,.
\ee
From a physical point of view the scattering length $a_0$ describes the low-energy scattering among particles, in the sense that
two particles interacting through the potential $V$, when they are far apart, feel the other particle as a hard sphere with radius $a_0$. 

The appearance of $a_0$ in the effective equation \eqref{GP} is a consequence of the fact that the many body Schr\"odinger evolution with Gross-Pitaevskii potential develops  a singular correlation structure which varies on the same length scale of the potential. Heuristically this can be seen considering the evolution equation for the one-particle reduced density matrix
\[ \label{hierarchy}
i \dpr_t \g^{(1)}_{N,t} = [-\D, \g^{(1)}_{N,t}] + \frac 1 {2} \Tr_2 \big[ V_N(x_1-x_2), \g^{(2)}_{N,t} \big]\,.
\]
To take into account correlations among the particles and the short scale structure they create in the marginal density $\g^{(2)}_{N,t}$, we may use the ansatz
\begin{equation} \label{ansatz}
\begin{split}
N^{-1} \g^{(1)}_{N,t}(x_1;x'_1) & =   \ph_t(x_1) \ph_t(x'_1)\,, \\
\binom{N}{2}^{-1}\g^{(2)}_{N,t}(x_1, x_2 ;x'_1, x'_2) & =  f_N(x_1 - x_2) f_N(x'_1 - x'_2) \ph_t(x_1) \ph_t(x_2) \ph_t(x'_1)\ph_t(x'_2) \,, 
\end{split}
\end{equation}
with $f_N(x)=f(Nx)$ the zero energy scattering function corresponding to the potential $N^2 V(Nx)$. Then Eq.~\eqref{GP} arises from \eqref{hierarchy} as the self consistent equation for $\ph_t$;  the coefficient in front of the non linearity is given by $\int \di x N^3 V(N(x)) f(Nx)=8\pi a_0$.
Note that the ansatz \eqref{ansatz} does not contradict complete condensation of the system at time t. On the contrary in the weak limit $N \to \io$ the function $f_N$ converges to one, and therefore $\g^{(2)}_{N,t}$ converges to 
$\ket{\ph_t} \bra{\ph_t}^{\otimes 2}$. 

This heuristics also explains why for $0<\b<1$ we get $\int V$ instead of $a_0$ in the effective equation for $\ph_t$, starting from the ansatz \eqref{ansatz}.   As shown for example in \cite[Lemma 2.1]{BCS} the potential $N^{3\b-1} V(N^\b x)$ has scattering length of order $N^{-1}$ for any choice of $0<\b<1$, 
and the solution $f_{N^\b}(x)$ of the scattering equation with potential $N^{3\b-1} V(N^\b x)$ satisfies the bound 
\[
\big(1 -f_{N^\b} \big)(x) \leq \frac{C}{N(|x| + N^{-\b})} \qquad   \text{for }  0<\b<1\,.
\]
Therefore the coefficient appearing  in front of the non linearity in the self consistent equation for $\ph_t$, obtained from \eqref{hierarchy} under the assumptions \eqref{ansatz} with $f_N(x)$ substituted by $f_{N^\b}(x)$, is
\[
 \int \di x \, N^{3\b} V(N^\b x) f_{N^\b}(x) = \int V  - c N^{\b-1}\,,
\]
which equals $\int V$ in the limit $N\to \io$, for any $0<\b<1$. Thus $\b=1$ is the only scaling for which the coefficient of the non linearity in the effective equation for $\ph_t$ is given by the scattering length of the unscaled potential $V$.

The rigourous derivation of the effective equation \eqref{Hartree} (the \emph{Hartree equation}), in the form presented in \eqref{tracenormconvergence},  has been first obtained by Spohn \cite{Spohn} for bounded potentials, and Erd\H{o}s-Yau \cite{EY, BGM} for singular potentials, analysing the BBGKY hierarchy for the density matrices. 
More recent approaches by Rodnianskii-Schlein \cite{RS} and Knowles-Pickl \cite{KP} also give the rate of convergence towards the Hartree dynamics.  
A derivation of the \emph{Gross-Pitaevskii equation} \eqref{GP} was obtained in a series of works \cite{ESY0, ESY2, ESY3} and later with an alternative approach in \cite{P}. More recently, convergence towards the Gross-Pitaevskii dynamics with a precise rate of convergence has been obtained in  \cite{BdOS12}.
The derivation of the non linear Schr\"odinger equation in the intermediate regimes $0<\b <1$  may be obtained with the same approaches, and it is in fact a simpler problem than the $\b=1$ case (see for example \cite{ESY1,P}; the proof in \cite{BdOS12} could be also easily adapted to cover any $\b<1$).

Beyond the approximation for the reduced density matrices, there is some interest in obtaining 
an approximation for the evolved $N$-particle wave function $\ps_{N,t}$ in the appropriate Hilbert space norm. This corresponds to study fluctuations around the effective dynamics described by the non linear Schr\"odinger equation for the condensate wave function.  Several results in this direction have been obtained in the mean field regime, starting from the pioneering works by Hepp and Ginibre-Velo \cite{Hepp,GV} and later in~\cite{GMM1, GMM2, XC, BKS, BSS, LNS, MPP}.
More recent results deal with the intermediate scalings with $\b>0$, see \cite{GM1, BCS, NN, NN2, GM2}. In particular, the result in \cite{BCS} covers all $\b<1$.  An analogous result for the Gross-Pitaevskii regime is up to now still open. 

More generally, from a statical point of view, one would aim to completely construct the ground state wave function and study its excitation spectrum at least in the interesting limiting regimes described by \eqref{HN} (and even more ambitiously in the thermodynamic limit). These goals have been partially achieved in the context of mean field bosons, where the ground state energy and excitation spectrum have been proved to be correctly described by the famous Bogoliubov approximation \cite{S, GS, DN, LNSS} and the ground state has been fully constructed in the presence of an ultraviolet cutoff \cite{Pizzo1, Pizzo2, Pizzo3}.  Up to now no similar results are available for any $\b>0$. From this point of view studying the fluctuation dynamics for $\b>0$ may also give some insight into the problem of approximating the ground state. 

The aim of this contribution is to present the  strategy used in \cite{BCS} to obtain a norm approximation for the dynamics described by the Hamiltonian $H_N^{(\b)}$, with $0<\b<1$. This approximation is obtained for  a special class of initial data in the Fock space. The choice of the initial data is a main point in our analysis, since in order to cover all $\b<1$ we need to introduce a suitable correlation structure among particles. We will come back to the role of correlations in our analysis in the next sections.

\section{The coherent state approach}  \label{sec:approach}

The strategy used in \cite{BCS}, also known as \emph{coherent state approach}, was first introduced by
Hepp in \cite{Hepp}. More recently it has been further developed in \cite{RS} and  \cite{BdOS12} to obtain the rate of the trace norm convergence in the mean field and Gross-Pitaevskii regimes respectively.  The main idea of this approach is that even if the dynamics described by $H_N^{(\b)}$ preserves the particle number, it is convenient to represent our bosonic system in the Fock space, where we have the opportunity to consider a more general class of initial data than wave functions in $L^2_s(\bR^{3N})$. The choice of the class of initial data crucially depends on the scaling of the potential. For this reason we first describe which choice turns out to be convenient in the mean field regime, and then present the physical and mathematical motivations leading to a different choice in the Gross-Pitaevskii regime. Before that, let us start with summarising the Fock space representation of a bosonic system.

\subparagraph{A. Fock space representation}

We represent our bosonic system  in  the bosonic Fock space
\be
{\cal F} = \oplus_{n \geq 0} L_s^2(\bR^{3n})\,.
\ee
A state $\Ps \in \cal F$ is therefore a sequence $\Psi = \{ \ps^{(n)}\}_{n \geq 0}$, where $\ps^{(0)} \in \bC$ and $\ps^{(n)} \in L^2_s(\bR^{3n})$. The space $\cF$ is a Hilbert space with respect to the inner product
\be
\bmedia{\Psi, \Phi} = \lis{\ps^{(0)}} \ph^{(0)} + \sum_{n \geq 1} \bmedia{\ps^{(n)}, \ph^{(n)}}\,,
\ee
and each component of $\Psi \in \cF$ has a probabilistic interpretation, namely $\|\ps^{(n)}\|^2_{2}$  is the probability of having $n$ particle in the state described by $\psi$. Clearly we are interested in states where $  \sum_{n\geq 0 }\|\ps^{(n)}\|^2_2 =1 $.  
The number of particles operator is defined requiring that
\be
({\cal N} \Psi)^{(n)} = n \ps^{(n)} \,,  \label{defN}
\ee
and therefore the expected number of particles in a state $\Psi \in \cF$ is given by 
\be
\bmedia{\Psi, \cN \Psi} = \sum_{n \geq 0} n \|\ps^{(n)}\|^2_{2}\,.
\ee
A state with exactly $N$ particles is represented by a vector in $\cF$ where only the $N$-th component is non zero. A special example of such a state is the vacuum state with $\O = \{1, 0,0, \ldots, 0\}$, describing a state with no particles.  More in general, given a one-particle operator $O^{(1)}$ the corresponding operator $d \G(O^{(1)})$ on $\cF$ (called {\it the second quantization} of $O^{(1)}$) is defined by the requirement that
\[ \label{SecondQuantO}
\big(d \G(O^{(1)}) \Ps \big)^{(n)} = \sum_{i=1}^n O^{(1)}_i \ps^{(n)}
\]
where $O^{(1)}_i$ denotes the operator acting on $\bR^{3n}$ as $O^{(1)}$ on the $i$-th particle and as the identity on the other $(n-1)$ particles. 

In order to define a time evolution on $\cF$ we introduce the Hamilton operator $\cH_N^{(\b)}$, which is defined through its action on vectors of $\cF$:
\begin{eqnarray}
(\cH_N^{(\b)} \Psi)^{(n)} &&= \big(H_N^{(\b)} \big)^{(n)} \ps^{(n)}\,, \nonumber \\
\big(H_N^{(\b)} \big)^{(n)}&&= \sum_{i=1}^n (-\D_{x_i}) + \sum_{i<j}^n N^{3\b -1}V( N^\b (x_i-x_j))\,.
\end{eqnarray}
By definition the operator $\cH_N^{(\b)} $ acts on states with a variable number of particles but leaves all sectors with fixed number of particles invariant. Note that the scaling parameter $N$ in $\cH_N^{(\b)}$ in general has nothing to do with the number of particles of the system (which is not fixed now). To recover the relevant scaling limits we are interested in, we will consider in the following the evolution of states with expected number of particle $N$.

Being the number of particles in $\cF$ not fixed, it is useful to introduce operators that create or annihilate a particle. For $f \in L^2(\bR^3)$ we define  the creation operator $a^*(f)$ and the annihilation operator $a(f)$ by
\begin{eqnarray}
&\big(  a^*(f) \Ps \big)^{(n)}(x_1, \ldots, x_n) & = \frac 1 {\sqrt n} \sum_{j=1}^n f(x_j) \ps^{(n-1)}(x_1, \ldots, \not{x_j}, \ldots, x_n) \quad n\geq 1\,, \qquad  \label{astar}\\
&\big(  a(f) \Ps \big)^{(n)}(x_1, \ldots, x_n) & = \sqrt{n+1} \int \di x \lis{f(x)} \ps^{(n+1)}(x, x_1, \ldots,  x_n)\qquad n \geq 0\,,  \qquad\label{a}
\end{eqnarray}
and we set $ (a^*(f) \Ps)^{(0)}:=0$. It is simply to check that $a^*(f)= (a(f))^*$, and that  the following commutation relations hold:
\[ \label{commutators}
[a(f), a^*(g)] = \bmedia{f,g}_{L^2}\,, && [a(f), a(g)]=[a^*(f), a^*(g)] =0\,.
\]
We have $a(f) \O = 0$. The action of $(a^*(f))^N$ on the vacuum generates a state with exactly $N$ particles with wave function $f$, that is 
\be
(\sqrt{N!})^{-1} (a^*(f))^N = \{0, \ldots, 0,\, f^{\otimes N},0, \ldots\}\,.
\ee
We also introduce operator valued distribution $a_x^*$ and $a_x$,  defined by
\be
a^*(f) = \int\di x  f(x)a^*_x\,, \quad \text{ and } \quad a(f) = \int \di x \lis{f(x)}a_x\,,
\ee
which formally creates or annihilates a particle in the point $x$. From \eqref{commutators} we have $[a_x, a^*_y] =\d_{x,y}$ and $[a_x, a_y] =[a^*_x, a^*_y]$=0. 

The second quantization of any (densely defined) self adjoint operator can be conveniently expressed by means of $a^*_x$ and $a_x$, see e.g. \cite[Sec. 3]{Heidelberg} and \cite[Sec. 1.3]{L}. The expressions for the particle number operator and the Hamiltonian are  
\be \label{cN}
\cN = \int \di x a^*_x a_x \,,
\ee
and
\be \label{cHN}
\cH_N^{(\b)} = \int \di x  \nabla_x a^*_x \nabla_x a_x +\frac 12 \int \di x \di y N^{3\b -1} V(N^\b( x-y)) a^*_x a^*_y a_x a_y
\ee
respectively. The r. h. s. of \eqref{cN} and \eqref{cHN} should be understood in the sense of forms; for example \eqref{cN} means that for any $\Psi, \Phi \in \cF$ we have $\bmedia{\Psi, \cN \Phi}= \int \di x \bmedia{a_x \Psi, a_x \Phi}$. 

Moreover, the kernel of the one-particle reduced density matrix $\g^{(1)}$ associated to the state $\Ps \in \cF$ can be expressed as
\be
\g^{(1)}(x;y) = \bmedia{\Psi, a^*_x a_y \Psi}\,.
\ee
The expression \eqref{cN} for $\cN$ suggests that, although creation and annihilation operators are unbounded operators, they can be bounded with respect to the square root of the number of particles operator, in the sense that
\begin{align} \label{bounds_a}
\|a(f) \Psi\| & \leq \|f\|_2 \|\cN^{1/2} \Psi\|  \nonumber \\
\|a^*(f) \Psi\|& \leq \|f\|_2 \|(\cN+1)^{1/2} \Psi\|
\end{align}
for all $f \in L^2(\bR^3)$, $\Psi \in \cF$. Moreover, given a bounded one particle operator $O^{(1)}$ on the $L^2(\bR^3)$ space, its second quantization $d\G(O^{(1)})$, although generally unbounded,  is bounded with respect to the number of particles operator: 
\be \label{OperatorBounds}
| \bmedia{\Ps,  d \G(O^{(1)})  \Ps} | \leq \|O^{(1)}\| \bmedia{\Ps, \cN \Ps}, \quad \Ps \in \cF\,.
\ee
Properties \eqref{bounds_a} and \eqref{OperatorBounds} will be essential for our analysis.

\subparagraph{B. Choice of the class of initial states in the Mean Field regime} 

Our goal is to study the time evolution under $\cH_N^{(\b)}$ of a suitable class of initial data in $\cF$  with expected number of particles $N$ and one-particle reduced density matrix $\g^{(1)}_{N,0}$ satisfying \eqref{BEC} for some $\ph \in L^2(\bR^3)$. In the mean field regime $\b=0$ a natural choice  is to consider as class of initial data the so called \emph{coherent states}.

A coherent state with wave function $f \in L^2(\bR^3)$ is a linear combination of states with all possible number of particles, all described by the same wave function $f$. Such a state is built acting on the vacuum with the so called {\it Weyl operator}
\be
W(f) = \exp(a^*(f) - a(f))\,,
\ee
thus obtaining
\be \label{coherentstate}
W(f) \O = e^{-\|f\|^2/2} \left \{ 1, f, \frac{f^{\otimes 2}}{\sqrt{2!}}, \ldots, \frac{f^{\otimes n}}{\sqrt{n!}}, \ldots \right\}\,.
\ee
The Weyl operator is a unitary operator on $\cF$ which acts on the annihilation and creation operators as follows:
\[  W^*(f)\, a_x\, W(f) & = a_x +f(x) \nonumber \\
 W^*(f)\, a^*_x\, W(f) &= a^*_x + \lis{f(x)}\,. \label{actionW3}
\] 
Since the expected particle number of the coherent state $W(f) \O$ is equal to
\be\bmedia{W(f) \O, \cN W(f) \O} = \|f\|^2_2\,,\ee
a coherent state with expected particle number $N$ is given by
\be \label{initialStateMF}
W(\sqrt{N} \ph) \O\,, \quad \|\ph \|_{2} =1\,.
\ee
Using (\ref{actionW3}) it is also simple to check that the kernel of the one-particle density associated to $W(\sqrt{N} \ph) \O$ is 
\be
\g^{(1)}_{N}(x,y)=\bmedia{W(\sqrt{N} \ph) \O , a^*_x a_y W(\sqrt{N} \ph) \O }= N \lis{\ph(x)} \ph(y)\,.
\ee
For this class of initial data the following theorem was proven in \cite{RS}, in the mean field regime.
\begin{theorem}  \label{th:MF}
Let $V$ be a measurable function, satisfying the operator inequality $V^2(x) \leq C(1-\D)$ for some $C>0$ and let $\ph \in H^1(\bR^3)$. Let $\g^{(1)}_{N,t}$ be the one-particle reduced density associated with 
\[
\Ps_{N,t}=e^{-i t\cH_N^{(0)}} W(\sqrt N \ph) \O\,.  \nonumber
\]
Then, there exist constants $D,k >0$ s.t. 
\[
\Tr\, \big|\g^{(1)}_{N,t} - N \ket{\ph_t} \bra{\ph_t}  \big| \leq D e^{k |t|} \nonumber
\]
 for all $t \in \bR$ and all $N \in \bN$, with $\ph_t$ satisfying \eqref{Hartree} with initial data $\ph_0=\ph$.
\end{theorem}
Note that the assumptions on $V$ in Theorem \ref{th:MF} include the Coulomb case $V(x)= \pm 1/|x|$. The strategy to prove Theorem \ref{th:MF} is to define a unitary operator $U_N(t)$ through the requirement: 
\be \label{assumptionMF}
 \Ps_{N,t}=e^{-i t  \cH_N^{(0)}} W(\sqrt N \ph) \O := W(\sqrt N \ph_t) U_N(t) \O\,.
\ee
Note that if $U_N(t)$ was the identity operator, than the evolution of $W(\sqrt{N}\ph)\Omega $ under the mean field Hamiltonian would be exactly a coherent state with evolved wave function $\ph_t$. In this sense the vector $U_N(t)\O$ is a fluctuation vector and 
\be
U_N(t) = W^*(\sqrt N \ph_t)e^{-i t  \cH_N^{(0)}} W(\sqrt N \ph)\,.
\ee
can be interpreted as a fluctuation dynamics. Using the definition \eqref{assumptionMF} we can write the kernel of the one particle reduced density matrix associated to the evolved state $\Ps_{N,t}$ as follows
\be
\g^{(1)}_{N,t}(x,y)=\bmedia{ U_N(t)\O , W^*(\sqrt{N} \ph)  a^*_x a_y W(\sqrt{N} \ph)  U_N(t)\O }\,.
\ee
For any compact one-particle observable $O^{(1)}$ on $L^2(\bR^3)$ one has
\[
\Tr\; O^{(1)} \Big(  \g^{(1)}_{N,t} - N \ket{\ph_t}\bra{\ph_t}\Big) =\; & \sqrt{N}\, \bmedia{U_N(t)\O, [\;a^*(O^{(1)} \ph_t) + a(O^{(1)} \ph_t)\;]  U_N(t)\O} \nonumber \\
& + \bmedia{U_N(t)\O,  d \G(O^{(1)}) U_N(t)\O}\,,
\]
with $d \G(O^{(1)}) $ defined in \eqref{SecondQuantO}.
Using \eqref{bounds_a} and \eqref{OperatorBounds} we have
\be \label{mainboundMF}
\Big|\Tr\; O^{(1)} \Big(  \g^{(1)}_{N,t} - N \ket{\ph_t}\bra{\ph_t}\Big)\Big| \leq  \sqrt{N}\; \bmedia{U_N(t)\O, (\cN+1)U_N(t)\O}  \,.
\ee
Since the space of trace class operators on $L^2(\bR^3)$, equipped with the trace norm, is the dual of the space of compact operators, equipped with the operator norm, the proof of Theorem \ref{th:MF} ends up with controlling the r.h.s. of \eqref{mainboundMF}. In particular, to get a bound on the rate of the convergence of the many body evolution towards the mean field dynamics proportional to $\sqrt N$ it is enough to show that the number of particles with respect to the fluctuation dynamics $U_N(t)$ grows uniformly in $N$. To this aim we compute
\be
i \dpr_t \bmedia{U_N(t)\O,  \cN U_N(t)\O} =  \bmedia{U_N(t)\O, [\cL^{(0)}_N(t), \cN] U_N(t)\O}\,,
\ee
with 
\be
\cL^{(0)}_N(t) =  (i\partial_t W^*(\sqrt N \ph_t)W(\sqrt{N} \ph_t)  +W^*(\sqrt{N} \ph_t) \cH_N^{(0)} W(\sqrt{N}\ph_t) \,.
\ee
the generator of the fluctuation dynamics $U_N(t)$. In contrast with the original Hamiltonian, $\cL_N^{(0)}(t)$ contains terms which do not commute with $\cN$. As a consequence, the expectation of $\cN$ is not preserved along the evolution of $U_N$, that is fluctuations are going to grow. However, under the assumption on the regularity of the potential stated in Theorem \ref{th:MF} it can be shown that 
\be
\pm [\cL^{(0)}_N(t), \cN] \leq    C   \big( \cN + 1\big) \label{NdotMF}\,.
\ee
Using a Gronwall lemma, we obtain that   $\bmedia{U_N(t)\O, (\cN+1)U_N(t)\O} $ is bounded uniformly in $N$. The fact that $\ph_t$ should satisfy the Hartree equation \eqref{Hartree} arises quite naturally, because this is the condition to be imposed in order to cancel some terms of order $\sqrt{N}$ in the generator which are linear in $a^*_x$ and $a_x$ and therefore do not commute with $\cN$.  Some more work is needed to get the (optimal) rate of convergence in Theorem \ref{th:MF} rather than the factor $\sqrt{N}$ in \eqref{mainboundMF}, but this issue is not relevant for the aim of this contribution.

\subparagraph{C. Choice of the class of initial states in the Gross-Pitaevskii regime} 

We consider now the Gross-Pitaevskii regime $\b=1$. To get the trace norm convergence result in this regime, the initial data \eqref{initialStateMF} has to be suitably modified to take into account correlations among particles, that play now a crucial role. In fact the Gross-Pitaevskii evolution develops singular correlations which are not captured by an approximation given in terms of coherent states.  

From the mathematical point of view this reflects into the fact that we cannot approximate the evolution of the class of coherent states \eqref{initialStateMF} under $\cH_N^{(GP)}:=\cH_N^{(1)}$ with a new coherent state with evolved wave function given by the Gross-Pitaevskii equation \eqref{GP}. If we defined the fluctuation dynamics
\be
 \tl U_N(t) =W^*(\sqrt N \ph_t) e^{-i t \cH_N^{(GP)}} W(\sqrt  N \ph)\,,
\ee
analogously to what was done in the mean field regime,  then the number of fluctuations  $ \bmedia{\tl U_N(t)\O,  \cN \tl U_N(t)\O}$ is going to grow with $N$. In fact the generator of the dynamics $\tl U_N(t)$ contains some linear and quadratic terms in the annihilation and creation operators whose commutator with $\cN$ cannot be bounded uniformly in $N$.

The idea used in \cite{BdOS12} to implement the appropriate correlation structure in the Fock space is to define the correlation kernel
\be \label{kernelGP}
k_t(x,y) = -N \o(N(x-y) \ph^N_t(x) \ph^N_t(y)\,,
\ee
with $\o(x)=1-f(x)$,  $f(x)$ the solution of the zero energy scattering equation \eqref{scattering},
and  $\ph_t^N$ the solution of the following modified Gross-Pitaevskii equation:\footnote{The choice of using the solution of the modified Gross-Pitaevskii equation \eqref{GPmod} rather than the solution of Eq.~\eqref{GP} is due to technical reasons; however note that in the limit $N \to \io$ the solution of \eqref{GPmod} approaches the solution of \eqref{GP}, as shown in \cite[Proposition 3.1]{BdOS12}. Despite the operators $k_t$ being N-dependent we do not put an extra N-index to keep the notation light.} 
\be  \label{GPmod}
i \dpr_t \ph^N_t = - \D \ph^N_t + \big(N^3 V(N \cdot) f(N \cdot) \star |\ph^N_t|^2 \big) \ph^N_t\,.
\ee
It is simple to check that the function $\o(x)$ satisfies the bound $N \o(Nx) \leq C (|x| + 1/N)^{-1} $ and $k_t$ is the kernel of an Hilbert-Schmidt operator.   In the following we identify the function $k_t \in L^2(\bR^3 \times \bR^3)$ with the operator having $k_t$ as its integral kernel. 
%
%
 Using $k_t$ we define a unitary operator $T(k_t)$ acting on the Fock space $\cF$ by
\be \label{BogT}
T(k_t) = e^{\frac 1 2\int \di x \di y (\,k_t(x,y)a^*_x a^*_y - \lis{k}_t(x,y) a_x a_y \,)}\,.
\ee
The action of $T(k_t)$ on the creation and annihiliation operators can be explicitly computed. For any $f \in L^2(\bR^3)$ we have (see \cite[Lemma 2.3]{BdOS12})
\[
{ T^*(k_t)\, a(f) T(k_t)} &= a(\cosh_{k_t}(f)) + a^*(\sinh_{k_t}(\bar f)) \nonumber \\
{ T^*(k_t)\, a^*(f) T(k_t)} &= a^*(\cosh_{k_t}(f)) + a(\sinh_{k_t}(\bar f)) \nonumber\,,
\]
where we used the notation $\cosh_{k_t}$ and $\sinh_{k_t}$ for the linear operators on $L^2(\bR^3)$ given by
\[
\cosh_{k_t} = \sum_{n\geq 0} \frac{1}{(2n)!} (k_t \lis{k}_t)^n\,, && \sinh_{k_t} = \sum_{n\geq 0} \frac{1}{(2n+1)!} (k_t \lis{k}_t)^n k_t\,,
\]
where products of $k_t$ and $\bar k_t$ have to be understood as products of operators.  
%
%
We now use the unitary operator $T(k_t)$ to approximate the correlation structure developed by the many-body evolution. To this aim, we consider the evolution of initial data having the form
\be \label{initialdataGP}
\Psi_{N,0} =W(\sqrt N \ph) T(k_0) \O\,.
\ee
Initial data given by Eq.~\eqref{initialdataGP}, known as \emph{squeezed coherent states}, are a natural class of initial data approximating the ground state of Bose-Einstein condensates trapped in a volume of order one.  In fact they have expected particle number $N + \|k_t\|_2$ (with $\|k_t\|$ of order one) and energy equal at leading order to ground state energy for trapped bosons in the Gross-Pitaevskii regime, see \cite[Appendix A]{Heidelberg}. 
From a physical point of view a good approximation for the ground state energy of a system of $N$ bosons is believed to be  of the form $\ph^{\otimes N} \prod_{i<j} f(N(x_i-x_j))$. Then,  the class of states $W(\sqrt N \ph) T(k_0) \O \in \cF$ captures some of the correlations  which are believed to truly appear in the ground state of dilute bosonic systems.  

The trace norm convergence result in the Gross-Pitaevskii regime is obtained  studying  the dynamics of states of the form $\Psi_{N,0}= W(\sqrt{N}\ph)T(k_0)\Omega$  under the Gross-Pitaevskii Hamiltonian $\cH_N^{(GP)}$. The fluctuation operator $\cU_N(t)$ is defined through the requirement that the many body evolution preserves the form of the initial data, up for the evolution of $\ph$ into $\ph^N_t$, that is
\be \label{evolution}
\Ps_{N,t}=e^{-i t \mathcal H_N^{(GP)}}W(\sqrt{N}\ph)T(k_0)\Omega :=  W(\sqrt{N}\ph_t^N)T(k_t) \cU_N(t) \O\,.
\ee
If $\cU_N(t)$ was the identity operator then the evolution of a  state of the form \eqref{initialdataGP} would be a  state of the same type with evolved condensate wave function  $\ph_t^N$ given by the modified Gross-Pitaevskii equation \eqref{GPmod}. In this sense $\cU_N(t) \O$ is a fluctuation vector and we refer to $\cU_N(t)$ as a fluctuation dynamics. Using the definition \eqref{evolution} we can write the kernel of the one particle reduced density matrix associated to the evolved state $\Ps_{N,t}$ as follows
\be
\g^{(1)}_{N,t}(x,y)=\bmedia{ \cU_N(t)\O , T^*(k_t)W^*(\sqrt{N} \ph^N_t)  a^*_x a_y W(\sqrt{N} \ph^N_t) T(k_t)  \cU_N(t)\O }\,,
\ee
with 
\be \label{FDyn}
 \cU_N(t)=T^*(k_t)W^*(\sqrt N \ph^N_t)e^{-i t \mathcal H_N^{(GP)}}W(\sqrt N \ph)T(k_0)\,.
\ee
The generator of the  fluctuation dynamics $\cU_N(t)$ is given by
\begin{eqnarray} \label{generatorGP}
 && \cL_N(t)=  (i\partial_tT^*(k_t))T(k_t)  \\
&& + T^*(k_t) \big[ \, \big(i\partial_t W^*(\sqrt N \ph^N_t)\big)\,W(\sqrt{N} \ph^N_t)  + W^*(\sqrt{N} \ph^N_t) \cH_N^{(GP)} W(\sqrt{N}\ph^N_t) \, \big] T(k_t)\,,  \nonumber
\end{eqnarray}
where 
\[
T^*(k_t) W^*(\sqrt{N} \ph^N_t) a_x W(\sqrt{N}\ph^N_t) T(k_t) &=\sqrt{N}\; \ph^N_t(x) + a(c_x) + a^*(s_x)  \nonumber \\
T^*(k_t) W^*(\sqrt{N} \ph^N_t) a^*_x W(\sqrt{N}\ph^N_t) T(k_t) &= 
\sqrt{N}\;  \lis{\ph^N_t(x)} + a^*(c_x) + a(s_x) \,.
\]
with $c_x(z)=\cosh_{k_t}(z, x)$ and $s_x(z)=\sinh_{k_t}(z, x)$.
Note that the action of  the Bogoliubov transformation $T(k_t)$ in Eq.~\eqref{generatorGP} generates terms in $ \cL_N(t) $ where the creation and annihilation operators are not in normal order (a product of creation and annihilation operators is said to be normal ordered if all creation operators are to the left of all annihilation operators). When we use the commutation relations \eqref{commutators} to restore the normal order, this procedure generates some new linear and quadratic terms in the creation and annihilation operators, coming from the normal ordering of 
some cubic and quartic terms respectively. 
These terms, together with the fact that the correlation kernel $k_t$ contains the solution $f_N$ of the scattering equation, lead to some cancellations in the generator $\cL_N$ which are essential to control the growth of the number of fluctuations uniformly in $N$. In particular, the sum of the linear terms  (which would be of order $\sqrt N$) gives zero when $\ph_t^N$ is chosen to satisfy the effective equation \eqref{GPmod}. A second cancellation arises between some quadratic terms that are too singular in the Gross-Piteavskii regime. After these cancellations (see \cite[Sect. 3]{BdOS12} for details) we have
\be
\pm [\cL_N(t), \cN] \leq  \cH_N + c \bestfrac{\cN^2}{N} +  C e^{k|t|}  \big( \cN + 1\big) \label{Ndot}
\ee
for some $C, c, k>0$ independent on $N$ and $t$. The time dependence on the r.h.s. of the last equation arises through high Sobolev norms of the solution $\ph_t$ of the Gross-Pitaevskii equation. 

The bound \eqref{Ndot} shows a further difference with respect to the strategy used to prove Theorem~\ref{th:MF}: in order to control the growth of the number of fluctuations $\bmedia{\cU_N(t)\O, \cN \cU_N(t) \O}$ in the Gross-Pitaevskii case we also need to control the growth of $\cH_N$. To this aim, in a very similar way as used to prove \eqref{Ndot} one can also obtain the bounds
\[ 
\cL_N(t) &\leq \frac 3 2\, \cH_N + c \bestfrac{\cN^2}{N}+ c e^{k|t|} (\cN+1)\,, \label{upperL}\\
\cL_N(t) &\geq \frac 1 2\, \cH_N - c \bestfrac{\cN^2}{N}- c e^{k|t|} (\cN+1)\,, \label{lowerL}\\
\pm \dot \cL_N(t) &\leq  2\, \cL_N(t) + c e^{k|t|}  ( \cN +1 + \bestfrac{\cN^2}{N}) \label{Ldot}\,.
\]
Moreover, it is easy to show that the number of fluctuations is just bounded by the total number of particles:
\[ \label{N2}
\bmedia{\cU_N(t) \O, (\bestfrac{\cN^2}{N}) \cU_N(t) \O} \leq \bmedia{\cU_N(t) \O, \cN \cU_N(t) \O}  + \bmedia{ \O,  (\bestfrac{\cN^2}{N})\,   \O}\,.
\]
Using \eqref{Ndot}, \eqref{upperL}, \eqref{Ldot} and \eqref{N2} one is able to close a Gronwall type estimate for the expectation 
\[ \label{Gronwall_GP}
\bmedia{\cU_N(t) \O, \big( \; \cL_N(t)  + D e^{k|t|} (\cN+1)\big) \cU_N(t) \O}\,,
\]
for some $D>0$, and show that it remains bounded uniformly in $N$.  We get finally the desired bound on the growth of $\cN$ observing that the lower bound \eqref{lowerL} together with \eqref{N2} implies
\[
\bmedia{U_N(t) \O, \big( \; 2\cL_N(t)  + c_1 e^{k|t|} (\cN+1)\big) \;U_N(t) \O} \geq \bmedia{U_N(t)\O, \cH_N U_N(t) \O} \geq 0\,, \nonumber
\]
for some $c_1>0$. Since $D$ can be chosen to be greater than $(c_1+1)$, the bound for \eqref{Gronwall_GP} also implies that  $\bmedia{U_N(t) \O, \cN U_N(t) \O}$ remains bounded uniformly in $N$. 
This allows to prove the following theorem, see \cite{BdOS12}.
\begin{theorem}\label{thm:GP}
Consider  a non-negative and spherically symmetric potential $V \in L^1 \cap L^3(\bR^3, (1+ |x|^6) \di x)$. Let $\ph \in H^4(\bR^3)$ and $\O \in \cF$ the vacuum state. Consider the family of initial data
\be
\Psi_N = W(\sqrt N \ph) T(k_0) \O \nonumber
\ee
and denote by $\g^{(1)}_{N,t}$ the one-particle reduced density matrix associated with the evolution $\Ps_{N,t}= e^{-i t\cH_N^{(GP)}} \Psi_N$. Then
\be
\Tr\, \big| \g^{(1)}_{N,t} - N \ket{\ph_t} \bra{\ph_t} \big| \leq C N^{1/2} \exp( \exp(c |t|)) \nonumber
\ee
for all $t \in \bR$. Here $\ph_t$ satisfies the Gross-Pitaevskii equation \eqref{GP} .
\end{theorem}
Theorem \ref{thm:GP} still holds if we substitute the vacuum state $\O$ with a sequence of states $\xi_N \in \cF$ such that $\|\x_N\|=1$ and 
$
\bmedia{\x_N, \big(\cH_N^{(GP)}+ \cN + \cN^2/N) \x_N} \leq C\,,
$
for some $C>0$ independent on $N$.

\section{Norm approximation result and ideas of the proof}

We switch now to the problem of studying fluctuations around the effective dynamics described by \eqref{Hartree}, \eqref{NLS} or \eqref{GP}.  The fact that the coherent state approach could also be used to describe fluctuations around the limiting equation has been first exploited in \cite{Hepp, GV, RS} in the mean field setting. 

In \cite{BCS} we follow the strategy used in \cite{RS}, the main difference coming from the necessity of taking into account correlations among particles in the condensate. As discussed in the introduction,  the many body evolution given by $\cH_N^{(\b)}$ for $0<\b<1$ develops weaker correlations than in the Gross-Pitaevskii regime. This is the reason why the effective dynamics is described by the non linear Schr\"odinger equation \eqref{NLS}, rather than the Gross-Pitaevskii equation \eqref{GP}. Anyway two body correlations are not negligible in the analysis of fluctuations. In fact, to get a norm approximation  result  valid for all $\b<1$,  we need to introduce a correlation structure, which is a suitable modification of the one defined in \eqref{kernelGP}. More precisely, instead of working with the kernel defined in \eqref{kernelGP} we consider 
\be \label{kell}
k_{\ell, t}(x;y) = - N \o_{N, \ell}(x-y) (\tl \ph^N_t((x+y)/2))^2\,.
\ee
Here $\tl \ph_t^N$ is the solution of the $N$-dependent Schr\"odinger equation
\be \label{nonlinear2}
i \dpr_t \tl\ph^N_t = - \D \tl\ph^N_t + \big(N^{3\b} V(N^\b \cdot) f_{N, \ell} \star |\tl\ph^N_t|^2 \big) \tl\ph^N_t\,,
\ee
 $\o_{N,\ell}= 1-f_{N,\ell}$, and $f_{N,\ell}$ is the solution of the eigenvalue problem
\be
 \Big[-\D + \frac 12 N^{3\b -1} V(N^\beta x) \Big] f_{N,\ell}(x) = \l_{N, \ell} f_{N,\ell}(x) \chi(|x| \leq \ell)\,,
\ee
associated with the smallest possible eigenvalue $\l_{N,\ell}$, normalized so that $f_{N,\ell}=1$ for $|x|=\ell$ and continued to $\bR^3$ by requiring that $f_{N,\ell}=1$ for all $|x|\geq \ell$. With this choice the kernel $k_{\ell,t}(x;y)=0$ for all $|x-y|> \ell$, that is we are considering particles correlated up to relative distance $\ell$.  
Note that, for all $0<\b<1$, the solution $\tl \ph_t^N$ of \eqref{nonlinear2}  approaches the solution of the non linear equation \eqref{NLS} as $N \to \io$.  However it furnishes a better approximation for the dynamics of the condensate wave function, since it contains the factor $f_{N, \ell}$ which takes into account the correlations among the particles. 

Using $k_{\ell,t}$ we define the Bogoliubov transformation $T(k_{\ell,t})$ through \eqref{BogT}. For any  $0<\b<1$ we consider the evolution of initial data of the form $W(\sqrt{N} \ph) T(k_{\ell,0}) \O$, defining the fluctuation dynamics:
\be  \label{Uell}
 \cU_{\ell,N}(t)=T^*(k_{\ell,t})W^*(\sqrt N \tl \ph^N_t)e^{-i t\mathcal H_N^{(\b)}}W(\sqrt N \ph)T(k_{\ell,0})\,, 
 \ee
 with $\tl \ph_0^N=\ph$.   The following result holds.

\begin{theorem} \label{thm:BCS}
Let $V\geq 0$, smooth, spherically symmetric and compactly supported. Fix $0<\b<1$ and consider $\cH_N^{(\b)}$ defined in \eqref{cHN}. Let $\tl \ph^N_t$ defined in \eqref{nonlinear2} with $\tl \ph_0^N= \ph \in H^4(\bR^3)$.  Fix $\ell>0$ and consider $k_{\ell,t}$  defined in \eqref{kell}. Let $\a= \min (\b/2, (1-\b)/2)$.  Then there exist  a unitary evolution  $\cU_{2,N}(t)$ with a quadratic (in the creation and annihilation operators) generator and constants $C, c_1, c_2>0$ such that 
\begin{eqnarray}
 \|  \; e^{-i t \cH_N^{(\b)}} W(\sqrt{N} \ph) T(k_{\ell,0}) \O \;-\; e^{-i \int_0^t \eta_N(s) \di s} && W(\sqrt{N} \tl \ph^N_t)\,  T(k_{\ell,t}) \, \cU_{2,N}(t) \O \; \|^2 \nonumber \\
 && \leq C\,N^{-\a}\, e^{c_1 \exp{c_2 |t|}}
 \end{eqnarray}
for all $t \in \bR$ and $N$ large enough. 
\end{theorem}

Theorem \ref{thm:BCS} still holds if we substitute the vacuum state $\O$ with a sequence of states $\xi_N \in \cF$ such that $\|\x_N\|=1$ and 
$
\bmedia{\x_N, \big(\cN^2 + \cK^2 + \cH_N^{(\b)} \big) \x_N} \leq C
$
uniformly in $N$. Here $\cK= \int \di x \nabla_x a ^*_x \nabla_x a_x$ is the kinetic energy operator.   It is also possible to approximate the dynamics of the fluctuations by a limiting evolution  $\cU_{2,\io}(t)$, again with a quadratic generator, but now independent of $N$, as shown in \cite[Prop. 2.1]{BCS}.    


While we refer to \cite{BCS} for a complete proof of Theorem \ref{thm:BCS}, we briefly describe here  the general strategy used there.  The main idea is to identify a limiting fluctuation dynamics with a quadratic generator, and then apply it to obtain the norm bound for the many body dynamics of our class of initial data. The fact that this limiting dynamics may exist is suggested by the form of the generator $\cL_{\ell,N}(t)$ of the dynamics $\cU_{\ell,N}$, where the cubic and quartic terms seem to vanish in the limit of large $N$.   From \eqref{Uell}  is apparent that, $W(f)$ and $T(k_{\ell,t})$ being  unitary operators, the following proposition is sufficient to prove Theorem \ref{thm:BCS}. 

\begin{proposition}\label{normProp} Let $\cU_{\ell,N}$ defined in \eqref{Uell}, and $\a= \min (\b/2, (1-\b)/2)$. Then, there exist  a unitary quadratic evolution $\cU_{2,N}$ and constants $C, c_1, c_2 >0$ such that,  for all $t \in \bR$ and all $N$ large enough,
\[
\|\cU_{\ell,N}(t;0) \O - e^{-i \int_0^t \eta_N(s) \di s} \cU_{2,N}(t;0) \O \| \leq C N^{-\a} \exp(\exp(c_2 |t|))\,.
\]
\end{proposition}
The proposition follows from the fact that the generator $\cL_{\ell,N}(t)$ can be written as
\be \label{generator-form}
\cL_{\ell,N}(t) = \eta_N(t)  + \cL_{2,N}(t) + \cV_N + \cE_N(t)\,,
\ee
where $\eta_N(t)$ is a phase,  $\cL_{2,N}(t)$ is a quadratic generator,
\be \cV_N = \frac 12 \int \di x \di y N^{3\b -1} V(N^\b( x-y)) a^*_x a^*_y a_x a_y
\ee  
is the interaction, and  $\cE_N(t)$  satisfies
 \[\label{eq:est-cE2} 
\left| \langle \psi_1, \cE_N (t) \psi_2 \rangle \right|  \leq  C N^{-\alpha} e^{K|t|}\big[ \langle \psi_1, (\cK &+ \cN + 1) \psi_1 \rangle \nonumber \\
&+ \langle \psi_2 , (\cK^2 + (\cN+1)^2) \psi_2 \rangle \big] \]
for all $\psi_1, \psi_2 \in \cF$. 
To prove Proposition \ref{normProp} we use that
\[ \label{GrowthU}
&\frac{\di}{\di t}\| \cU_{\ell,N}(t) \O - e^{-i \int_0^t \eta_N(s) \di s} \cU_{2,N}(t) \O\|^2 \nonumber \\
& \qquad = 2\Im \bmedia{\cU_{\ell,N} \O, (\cL_{\ell,N}(t) - \cL_{2,N}(t) - \eta_N(t)) e^{-i \int_0^t \eta_N(s) \di s} \cU_{2,N}(t) \O}\,,
\]
with $\cU_{2,N}(t) $ the dynamics generated by $\cL_{2,N}(t) $.  The r.h.s. of \eqref{GrowthU} is controlled using~\eqref{generator-form} and~\eqref{eq:est-cE2}: 
\[ \label{mainbound}
& | \bmedia{\cU_{\ell,N}(t) \O, (\cV_N + \cE_N(t)) \cU_{2,N}(t) \O}| 
\nonumber \\ & \hskip 1.5cm 
\leq C N^{-\a} e^{k|t|} \Big[\; \bmedia{\cU_{\ell,N}(t) \O, (\cH_N + \cN +1) \cU_{\ell,N}(t) \O} \nonumber \\
& \hskip 4 cm + \bmedia{\cU_{2,N}(t) \O, (\cK^2 + \cN^2 +1) \cU_{2,N}(t) \O}\; \Big]\,.
\]
The problem ends up in showing that the expectations appearing in the r.h.s. of  \eqref{mainbound} are all bounded uniformly in $N$. The  growth of $\cN$ and $\cH_N$ with respect to the full dynamics $\cU_{\ell,N}$ are controlled by means of Gronwall type estimates for the expectation of $\cN$ and  $\cL_{\ell,N}(t)$, following the same strategy described at the end  of Section~\ref{sec:approach} for the trace norm convergence result. The new issue here is that we also need to prove bounds for the growth of the expectation of $\cN^2$ and $\cK^2$ with respect to the dynamics generated by the quadratic part of the generator $\cL_{2,N} (t)$. We prove that the quadratic generator $\cL_{2,N} (t)$ satisfies the bounds 
\[ \label{eq:est-cL2N} 
\pm (\cL_{2,N} (t) - \cK) &\leq C e^{K|t|} (\cN+1), \qquad \hspace{.25cm} (\cL_{2,N} (t) - \cK)^2 \leq C e^{K|t|} (\cN+1)^2 \nonumber \\
\pm \left[ \cN , \cL_{2,N} (t) \right] &\leq C e^{K|t|} (\cN+1),  \qquad \hspace{.07cm} \pm \left[ \cN^2 , \cL_{2,N} (t) \right] \leq C e^{K|t|} (\cN+1)^2 \nonumber \\
\pm \dot{\cL}_{2,N} (t) &\leq C e^{K|t|} (\cN+1) ,\qquad 
\hspace{1.1cm} |\dot{\cL}_{2,N} (t)|^2 \leq C e^{K|t|} (\cN+1)^2\,. 
\]
Using Gronwall's Lemma and the bounds in~\eqref{eq:est-cL2N}, we obtain
\[\label{eq:bds-U2}  
\langle \cU_{2,N} (t;0) \O, \cN^2 \cU_{2,N} (t;0) \O\rangle &\leq C \exp (c_1 \exp (c_2 |t|)) \langle \O, (\cN+1)^2 \O \rangle  \nonumber \\
\langle \cU_{2,N} (t;0) \O, \cL_{2,N}^2 (t) \cU_{2,N} (t;0)\O \rangle &\leq C \exp (c_1 \exp (c_2 |t|)) \langle \O, (\cK + \cN +1)^2 \O\rangle\,.  \nonumber
\]
The last bounds, combined with the bound for $(\cL_{2,N} (t) - \cK)^2$, also implies that 
\be
\langle \cU_{2,N} (t;0) \O, \cK^2 \, \cU_{2,N} (t;0) \O \rangle \leq C \exp (c_1 \exp (c_2 |t|)) \langle \O, (\cK + \cN +1)^2 \O \rangle\,.  \nonumber
\ee
Note that some of the bounds in \eqref{eq:est-cL2N}  would not hold if we used the correlation structure defined in \eqref{kernelGP}; this is the reason why we implemented correlations through the kernel defined in \eqref{kell}. 

In \cite{BCS} we considered fluctuations around the non linear Schr\"odinger  dynamics for initial states on the Fock space. For $N$ particle initial data a more convenient approach to study fluctuations around the effective dynamics has been introduced in \cite{LNS} in mean field regime. This approach was later exploited in \cite{NN, NN2} to analyze fluctuations in the regimes up to $\b<1/2$. The major difficulty in the extension of these results to larger values of $\b$ is the introduction of correlations in the $N$ particle approach proposed in \cite{LNS}.

\section{Conclusions and open problems}

We reported on the proof of a norm approximation for the many-body dynamics described by  \eqref{cHN} of a particular class of initial data  in the Fock space which is a good candidate to describe the ground state of trapped bosons interacting with a pair potential of the form $N^{3\b-1} V(N^\b x)$, with $0<\b<1$.  In particular we showed that for any $0<\b<1$ one can approximate the fluctuation dynamics $\cU_{\ell,N}$ defined in \eqref{Uell} by a quadratic evolution in norm. 

Instead of considering fluctuations of the time evolution around the time dependent non linear Schr\"odinger equation, it is also possible to approach the problem from a static point of view. To this end, one can trap the system in a finite volume (either by imposing boundary conditions or by turning on an external potential) and one can study the difference between the many-body ground state energy and the minimum of the energy functional 
\[ \label{energyfunctional}
\cE(\ph) = \int \di x \big[ |\nabla \ph(x)|^2 + (\int V) |\ph(x)|^4 \big]\,.
\]
In this respect Theorem~\ref{thm:BCS} suggests that a good approximation for the many-body ground state of the Hamiltonian $\cH_N^{(\b)}$, with $0<\b<1$, may have the form $W(\sqrt{N} \ph) S \O$, where $\ph$
minimizes the energy functional  \eqref{energyfunctional} and $S$ is the exponential of a quadratic expression, related to the limiting quadratic evolution. 
Similarly, a good approximation for low-lying excited states may  be of the form $ W(\sqrt{N_0} \ph) S a^*(g_1) \ldots a^*(g_k) \O$,  for appropriate $k \in \bN$, $N_0=N-k$ and orbitals $g_1, \ldots, g_k$ orthogonal to $\ph$. It would be very interesting to obtain a proof of the above mentioned conjectures. 

Concerning the extension of our result to the Gross-Pitaevskii regime, new ideas are needed. In fact, if we follow the same strategy that we use for $\b<1$, it turns out that in the Gross-Pitaevskii regime one cannot approximate the fluctuation dynamics $\cU_{\ell,N}$ by a quadratic evolution in norm. In fact,  although one can control their effect on the growth of the number of particles (needed to prove the trace norm convergence), the cubic and quartic components of the generator of $\cU_{\ell,N}$ 
are not negligible in the limit of large $N$ as soon as $\b=1$.  In other words the fluctuation dynamics of quasi--free states is not described by a quadratic generator. One may interpret this difficulty saying that the action of $T(k_{\ell,0})$ is not sufficient to describe the correlation structure developed in the Gross-Pitaevskii regime with the precision needed to get a norm approximation result.  
In this perspective the analysis of the fluctuation dynamics around the Gross-Pitaevskii equation may be useful to get some information on the ground state wave function in this physically relevant regime. Vice versa some new results on the time-independent characterization of bosonic systems in the Gross-Pitaevskii regime may help the understanding of the dynamical properties of the system.

We hope to be able to address some of these problems in the next future.

%
%
%
%
%


\end{document}